\documentclass[11pt,a4paper]{article}
\usepackage{emnlp2019}
\usepackage{times}
\usepackage{graphicx}
\usepackage{booktabs}

\aclfinalcopy 

\newcommand*\samethanks[1][\value{footnote}]{\footnotemark[#1]}

\newcommand{\ignore}[1]{}

\title{Simple Applications of BERT for Ad Hoc Document Retrieval}

\author{Wei Yang,\thanks{\hspace{0.25cm}equal contribution} \hspace{0.1cm} Haotian Zhang,\samethanks \hspace{0.1cm} \and Jimmy Lin\vspace{0.1cm}\\
  David R. Cheriton School of Computer Science \\
  University of Waterloo}

\date{}

\begin{document}
\maketitle
\begin{abstract}
Following recent successes in applying BERT to question answering, we explore simple applications to {\it ad hoc} document retrieval.
This required confronting the challenge posed by documents that are typically longer than the length of input BERT was designed to handle.
We address this issue by applying inference on sentences individually, and then aggregating sentence scores to produce document scores.
Experiments on TREC microblog and newswire test collections show that our approach is simple yet effective, as we report the highest average precision on these datasets by neural approaches that we are aware of.
\end{abstract}

\section{Introduction}

The dominant approach to {\it ad hoc} document retrieval using neural networks today is to deploy the neural model as a reranker over an initial list of candidate documents retrieved using a standard bag-of-words term-matching technique.
Researchers have proposed many neural ranking models~\cite{MitraBhaskar_Craswell_2019}, but there has recently been some skepticism about whether they have truly advanced the state of the art~\cite{Lin_SIGIRForum2018}, at least in the absence of large amounts of log data only available to a few organizations.

One important recent innovation is the use of neural models that make heavy use of pretraining \cite{N18-1202,radford2018improving}, culminating in BERT \cite{devlin2018bert}, the most popular example of this approach today.
Researchers have applied BERT to a broad range of NLP tasks and reported impressive gains.
Most relevant to document retrieval, BERTserini~\cite{Yang_etal_arXiv2019} integrates passage retrieval using the open-source Anserini IR toolkit with a BERT-based reader to achieve large gains over the previous state of the art in identifying answer spans from a large Wikipedia corpus.

Given the successes in applying BERT to question answering and the similarities between QA and document retrieval, we naturally wondered:\ Would it be possible to apply BERT to improve document retrieval as well?
In short, the answer is {\it yes}.
Adapting BERT for document retrieval requires overcoming the challenges associated with long documents, both during training and inference.
We present a simple yet effective approach, based on the same BERTserini framework, that applies inference over individual sentences in a document and then combines sentence scores into document scores.

Our approach is evaluated on standard {\it ad hoc} retrieval test collections from the TREC Microblog Tracks (2011--2014) and the TREC 2004 Robust Track.
We report the highest average precision on these datasets for neural approaches that we are aware of.
The contribution of our work is, to our knowledge, the first successful application of BERT to {\it ad hoc} document retrieval, yielding state of the art results.

\section{Background and Related Work}

In {\it ad hoc} document retrieval, the system is given a short query $q$ and the task is to produce the best ranking of documents in a corpus, according to some standard metric such as average precision (AP).
\citet{MitraBhaskar_Craswell_2019} provide a recent overview of many of these models, to which we refer interested readers in lieu of a detailed literature review due to space considerations.

However, there are aspects of the task worth discussing.
Researchers have understood for a few years now that {\it relevance matching} and {\it semantic matching} (for example, paraphrase detection, natural language inference, etc.)~are different tasks, despite shared common characteristics~\cite{Guo:2016:DRM:2983323.2983769}.
The first task has a heavier dependence on exact match (i.e., ``one-hot'') signals, whereas the second task generally requires models to more accurately capture semantics.
Question answering has elements of both, but nevertheless remains a different task from document retrieval.
Due to these task differences, neural models for document ranking, for example, DRMM~\cite{Guo:2016:DRM:2983323.2983769}, are quite different architecturally from neural models for capturing similarity; see, for example, the survey of~\citet{C18-1328}.

Another salient fact is that documents can be longer than the length of input texts that BERT was designed for.
This creates a problem during training because relevance judgments are annotations on documents, not on individual sentences or passages.
Typically, within a relevant document, only a few passages are relevant, but such fine-grained annotations are not available in most test collections.
Thus, it is unclear how exactly one would fine-tune BERT given (only) existing document-level relevance judgments.
In this paper, we sidestep the training challenge completely and present a simple approach to aggregating sentence-level scores during inference.

\begin{table*}[t]
\centering\resizebox{\textwidth}{!}{
\begin{tabular}{ll ll ll ll ll }
\toprule
 & \multicolumn{2}{l}{\textbf{2011}} & \multicolumn{2}{l}{\textbf{2012}} & \multicolumn{2}{l}{\textbf{2013}} & \multicolumn{2}{l}{\textbf{2014}} \\
{\bf Model} & {\bf AP} & {\bf P30} & {\bf AP} & {\bf P30} & {\bf AP} & {\bf P30} &  {\bf AP} & {\bf P30} \\
\toprule
QL                             & 0.3576 & 0.4000 & 0.2091 & 0.3311 & 0.2532 & 0.4450 & 0.3924 & 0.6182 \\
RM3                            & 0.3824 & 0.4211 & 0.2342 & 0.3452 & 0.2766 & 0.4733 & 0.4480 & 0.6339 \\
\hline
DRMM~\cite{Guo:2016:DRM:2983323.2983769}    & 0.3477 & 0.4034 & 0.2213 & 0.3537 & 0.2639 & 0.4772 & 0.4042 & 0.6139 \\
DUET~\cite{Mitra:2017:LMU:3038912.3052579}  & 0.3576 & 0.4000 & 0.2243 & 0.3644 & 0.2779 & 0.4878 & 0.4219 & 0.6467 \\  
K-NRM~\cite{Xiong:2017:ENA:3077136.3080809} & 0.3576 & 0.4000 & 0.2277 & 0.3520 & 0.2721 & 0.4756 & 0.4137 & 0.6358 \\
PACRR~\cite{D17-1110}                       & 0.3810 & 0.4286 & 0.2311 & 0.3576 & 0.2803 & 0.4944 & 0.4140 & 0.6358 \\
MP-HCNN~\cite{rao2019multi}                 & 0.4043 & 0.4293 & 0.2460 & 0.3791 & 0.2896 & 0.5294 & 0.4420 & 0.6394 \\
\midrule
BiCNN \cite{Shi_etal_arXiv2018}             & 0.4293 & 0.4728 & 0.2621 & 0.4147 & 0.2990 & 0.5367 & 0.4563 & 0.6806 \\
\midrule
{BERT} & \textbf{0.4697} & \textbf{0.5040} & \textbf{0.3073} & \textbf{0.4356} & \textbf{0.3357} & \textbf{0.5656} & \textbf{0.5176} & \textbf{0.7006} \\
\bottomrule
\end{tabular}
}
\caption{Results on test collections from the TREC Microblog Tracks, comparing BERT with selected neural ranking models. The first two blocks of the table contain results copied from \citet{rao2019multi}.}
\label{table:microblog}
\end{table*}

\section{Searching Social Media Posts}

Despite the task mismatch between QA and {\it ad hoc} document retrieval, our working hypothesis is that BERT can be fine-tuned to capture relevance matching, as long as we can provide appropriate training data.
To begin, we tackled microblog retrieval----searching short social media posts---where document length does not pose an issue.
Fortunately, test collections from the TREC Microblog Tracks~\cite{Lin_etal_TREC2014}, from 2011 to 2014, provide data for exactly this task.

As with BERTserini, we adopted a simple architecture that uses the Anserini IR toolkit\footnote{http://anserini.io/} for initial retrieval, followed by inference using a BERT model.
Building on best practice, query likelihood (QL) with RM3 relevance feedback~\cite{Abdul-Jaleel04} provides the initial ranking to depth 1000.
The texts of the retrieved documents (posts) are then fed into a BERT classifier, and the BERT scores are combined with the retrieval scores via linear interpolation.
We used the BERT-Base model (uncased, 12-layer, 768-hidden, 12-heads, 110M parameters) described in \citet{devlin2018bert}.
As input, we concatenated the query $Q$ and the document $D$ into a text sequence [[CLS], $Q$, [SEP], $D$, [SEP]], and then padded each text sequence in a mini-batch to $N$ tokens, where $N$ is the maximum length in the batch.
Following \citet{nogueira2019passage}, BERT is used for binary classification (i.e., relevance) by taking the [CLS] vector as input to a single layer neural network.

Test collections from the TREC Microblog Tracks were used for fine-tuning the BERT model, using cross-entropy loss.
For evaluation on each year's dataset, we used the remaining years for fine tuning, e.g., tuning on 2011--2013 data, testing on 2014 data.
From the training data, we sampled 10\% for validation.
We fine-tuned BERT with a learning rate of $3 \times 10^{-6}$ for 10 epochs.
The interpolation weight between the BERT scores and the retrieval scores was tuned on the validation data.
We only used as training examples the social media posts that appear in our initial ranking (i.e., as opposed to all available relevance judgments).
There are a total of 225 topics (50, 60, 60, 55) in the four datasets, which yields 225,000 examples (unjudged posts are treated as not relevant).

Experimental results are shown in Table~\ref{table:microblog}, where we present average precision (AP) and precision at rank 30 (P30), the two official metrics of the evaluation~\cite{Ounis_etal_TREC2011}.
The first two blocks of the table are copied from~\citet{rao2019multi}, who compared bag-of-words baselines (QL and RM3) to several popular neural ranking models as well as MP-HCNN, the model they introduced.
Results for all the neural models include interpolation with the original document scores.
\citet{rao2019multi} demonstrated that previous neural models are not suitable for ranking short social media posts, and are no better than the RM3 baseline in many cases.
In contrast, MP-HCNN was explicitly designed with characteristics of tweets in mind:\ it significantly outperforms previous neural ranking models (see original paper for comparisons, not repeated here).
We also copied results from \citet{Shi_etal_arXiv2018}, who reported even higher effectiveness than MP-HCNN.

These results represent, to our knowledge, the most comprehensive summary of search effectiveness measured on the TREC Microblog datasets.
Note that for these comparisons we leave aside many non-neural approaches that take advantage of learning-to-ranking techniques over manually-engineered features, as we do not believe they form a fair basis of comparison.
In general, such approaches also take advantage of non-textual features (e.g., social signals), and these additional signals (naturally) allow them to beat approaches that use only the text of the social media posts (like all the models discussed here).

The final row of Table~\ref{table:microblog} reports results using our simple BERT-based technique, showing quite substantial and consistent improvements over previous results.
Since we have directly copied results from previous papers, we did not conduct significance tests.

\section{Searching Newswire Articles}

Results on the microblog test collections confirm our working hypothesis that BERT can be fine-tuned to capture document relevance, at least for short social media posts.
In other words, task differences between QA and document retrieval do not appear to hinder BERT's adaptability.
Having demonstrated this, we turn our attention to longer documents.
For this, we take advantage of the test collection from the TREC 2004 Robust Track~\cite{Voorhees_TREC2004_robust}, which comprises 250 topics over a newswire corpus.
We selected this collection for a couple of reasons:\ it is the largest newswire collection we know of in terms of training data, and \citet{Lin_SIGIRForum2018} provides well-tuned baselines that support fair comparisons to recent neural ranking models.

Given the success of BERT on microblogs, one simple idea is to apply inference over each sentence in a candidate document, select the one with the highest score, and then combine that with the original document score (with linear interpolation).
One rationale for this approach comes from \citet{dblp:journals/corr/abs-1803-08988,Zhang:2018:EUI:3269206.3271796}, who found that the ``best'' sentence or paragraph in a document provides a good proxy for document relevance.
This is also consistent with a long thread of work in information retrieval that leverages passage retrieval techniques for document ranking~\citep{Callan_SIGIR1994,Clarke00a,liu2002passage}. 

Generalizing, we could consider the top $n$ scoring sentences as follows:
\[\textrm{Score}_d = a \cdot S_{\textrm{doc}} + (1-a) \cdot \sum_{i=1}^{n} w_i\cdot S_i \]
\noindent where $S_{\textrm{doc}}$ is the original document score and $S_i$ is the $i$-th top scoring sentence according to BERT.
The hyperparameters $a$ and $w_i$ can be tuned via cross-validation.

Sentence-level inference seems like a reasonable initial attempt at adapting BERT to document retrieval, but what about fine-tuning?
As previously discussed, the issue is that we lack sentence-level relevance judgments.
Since our efforts represent an initial exploration, we simply sidestep this challenge (for now) and fine tune on existing sentence-level datasets.
Specifically, we used:\
(1) the microblog data from the previous section and (2) the union of the TrecQA~\cite{Yao13answerextraction} and Wiki\-QA~\cite{yang2015wikiqa} datasets.
This sets up an interesting contrast:\ the first dataset captures the document retrieval task but on a different domain, while the second dataset captures a different task but on corpora that are much closer to newswire.
It is an empirical question as to which source is more effective.

To support a fair comparison, we adopted the same experimental procedure as \citet{Lin_SIGIRForum2018}.
He described two separate data conditions:\ one based on two-fold cross-validation to compare against ``Paper 1'' and one based on five-fold cross-validation to compare against ``Paper 2''.\footnote{Since Lin's article is critical of neural methods, he anonymized the neural approaches but mentioned that they come from articles published in late 2018 and are representative of the most recent advances in neural approaches to document retrieval.}
The exact fold settings are provided online, which ensures a fair comparison.\footnote{https://github.com/castorini/Anserini/blob/master/docs/ experiments-forum2018.md}
In our implementation, documents are first cleaned by stripping all tags and then segmenting the text into sentences using NLTK.
If the input to BERT is longer than 512 tokens (BERT's maximum limit), we further split sentences into fixed sized chunks.
Across the 250 topics, each document averages 43 sentences, with 27 tokens per sentence.

In our experiments, we considered up to the top four sentences.
For up to three sentences, $a$ and $w_i$ are tuned via exhaustive grid search in the following range:\
$a\in [0, 1]$, $w_1=1$ (fixed), $w_2\in [0, 1]$, and $w_3\in [0, 1]$, all with step size $0.1$.
In the four-sentence condition, to reduce the search space, we started with the best three-sentence parameters and explored $w_4\in [0, 1]$ 
with step size $0.1$, along with neighboring regions in $a$, $w_2$, and $w_3$.
We selected the parameters with the highest AP score on the training folds.

\begin{table}[t]
\centering\begin{tabular}{lll}
\toprule
Model & AP & P20 \\
\toprule
Paper 1 (two fold) & 0.2971 & 0.3948 \\
BM25+RM3 & 0.2987  & 0.3871 \\
1S: BERT FT(QA)        & 0.3014 & 0.3928 \\
2S: BERT FT(QA)        & 0.3003 & 0.3948 \\
3S: BERT FT(QA)        & 0.3003 & 0.3948 \\
1S: BERT FT(Microblog) & 0.3241 & 0.4217 \\
2S: BERT FT(Microblog) & 0.3240 & 0.4209 \\
3S: BERT FT(Microblog) & \textbf{0.3244} & \textbf{0.4219} \\
\midrule
Paper 2 (five fold) & 0.272 &  0.386 \\
BM25+RM3 & 0.3033 & 0.3974 \\
1S: BERT FT(QA) & 0.3102 & 0.4068 \\
2S: BERT FT(QA) & 0.3090 & 0.4064 \\
3S: BERT FT(QA) & 0.3090 & 0.4064 \\
1S: BERT FT(Microblog) & 0.3266 & 0.4245 \\
2S: BERT FT(Microblog) & \textbf{0.3278} & 0.4267 \\
3S: BERT FT(Microblog) & \textbf{0.3278} & \textbf{0.4287} \\
\bottomrule
\end{tabular}
\caption{Results on Robust04. FT indicates the dataset used for fine tuning; $n$S indicates inference using the top $n$ scoring sentences of the document.}
\label{table:newswire}
\end{table}

Results of our experiments are shown in Table~\ref{table:newswire}, divided into two blocks:\ Paper 1 on the top and Paper 2 on the bottom.
The effectiveness of the two papers are directly copied from \citet{Lin_SIGIRForum2018}; all other results are our own runs.
The paper aggregation site ``Papers With Code'' places Lin's result as the state of the art on Robust04 as of this writing.\footnote{https://paperswithcode.com/sota/\\ ad-hoc-information-retrieval-trec-robust}
As a point of comparison, in the most recent survey of neural ranking models by~\citet{Guo:1903.06902v1:2019}, the best AP on Robust04 is in the $0.29$ range, consistent with the above site.
Therefore, we are quite confident that we are evaluating against competitive models.
In the results table, ``FT'' indicates the dataset used for fine-tuning and $n$S indicates inference using the top $n$ scoring sentences of the document.
We find that the learned $w_4$ value is zero, indicating that additional sentences do not help beyond the top three (at least according to our tuning procedure); thus, 4S results are omitted from the table.
Interestingly, we find that fine-tuning BERT on microblog data is more effective than QA data, suggesting that task (QA vs.\ relevance matching) is more important than document genre (tweets vs.\ newswire).
Cognizant of the potential dangers of repeated hypothesis testing, we probed the statistical significance of one five-fold setting, BM25+RM3 vs.\ ``3S: BERT FT (Microblog)''.
According to a paired $t$-test, the differences are statistically significant ($p < 10^{-7}$).

As a summary, we see that a well-tuned BM25+RM3 baseline already outperforms neural ranking approaches (which was Lin's original point).
Our simple BERT-based reranker yields further significant improvements.


\section{Conclusions}

In this preliminary study, we have adapted BERT for document retrieval in the most obvious manner, via sentence-level inference and simple score aggregation.
Results show substantial improvements in both ranking social media posts and newswire documents---to our knowledge, the highest AP scores reported on the TREC Microblog and Robust04 datasets for neural approaches that we are aware of (although the literature does report non-neural approaches that are even better, for both tasks).
We readily concede that our techniques are quite simple and that there are many obvious next steps.
In particular, we simply sidestepped the issue of not having sentence-level relevance judgments, although there are some obvious distant supervision techniques to ``project'' relevance labels down to the sentence level that should be explored.
We are actively pursuing these and other directions.

\section*{Acknowledgments}

This research was supported by the Natural Sciences and Engineering Research Council (NSERC) of Canada. 


\end{document}